\documentclass[amssymb]{revtex4}

\usepackage{graphicx}

\newcommand{\br}{ {\bf r} }

\begin{document}
\title
{
Quantum phase transition in space in a ferromagnetic spin-1 Bose-Einstein condensate
}
\begin{abstract}
A quantum phase transition between the symmetric (polar) phase and
the phase with broken symmetry can be induced in a ferromagnetic 
spin-1 Bose-Einstein condensate in {\it space} (rather than in time).
We consider such a phase transition and show that the transition 
region in the vicinity of the critical point exhibits scalings 
that reflect a compromise between the rate at which the transition 
is imposed (i.e., the gradient of the control parameter) 
and the scaling of the divergent healing length in the critical region.
Our results suggest a method for the direct measurement of the
scaling exponent $\nu$.
\end{abstract}
\author{Bogdan Damski and Wojciech H. Zurek}
\affiliation
{
Theoretical Division, Los Alamos National Laboratory, MS-B213, Los Alamos, NM 87545, USA
}
\maketitle

\section{Introduction}

Studies of phase transitions have  traditionally focused 
on {\it equilibrium} scalings of various properties near the critical point
of a {\it homogeneous} system.
The dynamics of  phase transitions presents new interdisciplinary challenges.
Nonequilibrium  phase 
transitions may play a role in  the evolution of the early Universe \cite{kibble}.
Their analogues can be studied in  condensed matter
systems \cite{zurek}. The latter observation led to 
the series of beautiful experiments \cite{eksperymenty} (see
\cite{kibble_today} for an up-to-date  review) and to 
the development of the theory based on the universality of  critical
behavior \cite{zurek}.
The recent progress in cold atom experiments allows for the
temporal and spatial control  of different systems undergoing a quantum phase
transition (QPT) \cite{lewenstein,nature_kurn}. 
These experimental developments call 
for an in-depth  understanding of non-equilibrium QPTs.

A QPT is a fundamental change in the {\it ground state} of the system  as a result of
 small variations of an external parameter, e.g., 
 a magnetic field \cite{sachdev}.
In contrast to thermodynamic phase transitions, it takes place  ideally at 
temperature of absolute zero.

The problem of how a quantum system undergoes a transition 
from one quantum phase to another due to time-dependent 
(temporal) driving has attracted lots of attention lately
\cite{dorner,spiny1,spiny2,ferro_bodzio_prl,ferro_bodzio_njp,austen_and_ueda_1_6,spins_sen,spins_fazio}. 
Basic insights into the QPT dynamics can be obtained through
the quantum version \cite{dorner,lz_bodzio} of the Kibble-Zurek mechanism (KZM) 
\cite{kibble,zurek}. The KZM recognizes that the time 
evolution of the quantum  system is adiabatic far away from the
critical point where the gap in the excitation spectrum is large.
The system  adjusts to
all changes imposed on its Hamiltonian. Near the critical
point, however, the gap closes precluding the adiabatic evolution 
and resulting in the non-equilibrium dynamics. This switch of behavior 
happens when   
the system reaction time $\hbar/\Delta$ ($\Delta$ is excitation gap)
becomes comparable to the time
scale on which the critical point is approached, 
$\varepsilon/|d\varepsilon/dt|$ 
($\varepsilon=|q-q_c|$, $q$ is the parameter driving the transition,
$q_c$ is the location of the critical point). This brings us to 
\begin{equation}
\label{gap_eq}
\frac{\hbar}{\Delta} = \frac{\varepsilon}{|d\varepsilon/dt|}.
\end{equation}
To solve (\ref{gap_eq}) we define   
$\left|\frac{d}{dt}q(t)\right|=\tau_Q^{-1}$.
The solution of (\ref{gap_eq}) gives us time $\hat t$, left to 
reaching the critical point, when the non-equilibrium 
dynamics starts 
\begin{equation}
\label{t_hat}
\hat t= (\hbar/\Delta_0)^{1/(1+z\nu)}\tau_Q^{z\nu/(1+z\nu)}.
\end{equation}
Above we have set $\Delta=\Delta_0|q_c-q|^{z\nu}$ near the critical point
($z$ and $\nu$ are critical exponents).

The quantum KZM 
proposes that  the non-equilibrium evolution in the neighborhood 
of the critical
point is to a first approximation impulse (diabatic):
the state of the system does not change there. 
Suppose that the system evolves towards the critical
point  from the ground state far away from it. Thus its properties near  the
critical point are determined by its ground state properties at a distance 
$\hat t/\tau_Q$ from  the critical point 
(the adiabatic regime was abandoned there).
As have been shown  in spin models (e.g., \cite{dorner,spiny1})
and ferromagnetic spin-1 
Bose-Einstein condensates
\cite{ferro_bodzio_prl,ferro_bodzio_njp,austen_and_ueda_1_6}, 
this simplification allows for a correct 
analytical computation  of the density of excitations 
resulting from the non-equilibrium quench.

In this paper we study a quantum phase transition induced by  {\it
spacial}  rather than  temporal driving.
It means that 
the driving parameter, $q$, depends on position and is independent of time. 
Such a transition was recently studied in the quantum Ising model
\cite{platini,xzurek} and the mean-field Ginzburg-Landau theory \cite{platini}.
Here we present the theory of the spatial quench in a ferromagnetic spin-1
Bose-Einstein condensate, which is one of the most flexible
physical systems for studies of QPTs. 

We focus on the ground state. 
In the simplest approximation, the local homogeneous approximation (LHA), 
the system is locally characterized by its homogeneous ground state
properties perfectly tracking spatial variations of $q$.
This approximation is good far enough  from the critical 
point, where the healing length \cite{healing} is small compared to 
the imposed scale of spatial driving.  
Using  the time quench analogy, the system  evolves perfectly 
adiabatically in space away from the spatial critical point, $\br_c$, where $q(\br_c)=q_c$. 

Around the critical point, however, the healing length diverges
and so the LHA breaks down.
The system enters ``non-equilibrium'' regime similarly
as during a time quench.
The spatial coupling  term in the Hamiltonian
(a gradient term in mean-field theories,
a spin-spin interaction term in spin models)
smoothes out the sharp spatial boundary between the phases
predicted by the LHA.

Even more interestingly, these similarities are not only qualitative
\cite{xzurek}.
The  distance from the spatial critical point where the switch between the two regimes
takes place, $\hat x$, can be determined by looking at the length scales
relevant for the spatial quench. We compare  the healing length
\begin{equation}
\frac{\xi_0}{|q-q_c|^\nu},
\label{heal}
\end{equation}
to the length scale on which the spatial critical 
point is approached: $\varepsilon/|d\varepsilon/dx|$ assuming that  $\varepsilon=|q-q_c|$ changes in 
$x$-direction only. This brings us to the spatial analog of
(\ref{gap_eq})
\begin{equation}
\frac{\xi_0}{|q-q_c|^\nu} = \frac{\varepsilon}{|d\varepsilon/dx|}.
\label{gapx_eq}
\end{equation}
To solve (\ref{gapx_eq}) we define  $\left|\frac{d}{dx}q(x)\right|=\lambda_Q^{-1}$ 
and get 
\begin{equation}
\hat x= \xi_0^{1/(1+\nu)}\lambda_Q^{\nu/(1+\nu)}.
\label{x_hat}
\end{equation}
The results (\ref{t_hat}) and (\ref{x_hat}) are 
analogous. Indeed, apart from a slight difference in the 
scaling exponents -- due to the absence of $z$ in (\ref{x_hat}) -- 
$\hat t$ maps onto $\hat x$ when time
scales $\hbar/\Delta_0$ and $\tau_Q$ 
are replaced by length scales $\xi_0$ and  $\lambda_Q$, respectively.
This shows striking parallels between quenches in time and space.
It is also instructive to note that the same general result, Eq.
(\ref{x_hat}), can be
obtained in a different way from the scaling theory \cite{platini}.

\section{Model}

In the following, we will study a QPT in space in 
a ferromagnetic spin-1 Bose-Einstein condensate.
We consider  untrapped clouds: atoms in a box. Such a system 
can be realized in an optical box trap \cite{raizen}.
Assuming that the condensate is placed in the 
magnetic field $B(\br)$ aligned in the $z$ direction, its 
mean-field energy functional reads \cite{leshouches}
\begin{equation}
\label{energy}
{\cal E}[\Psi] = \int d{\bf r} \ 
         \frac{\hbar^2}{2M}|\vec{\nabla}\Psi|^2
       + \frac{c_0}{2}\langle\Psi|\Psi\rangle^2 
       + \frac{c_1}{2}\sum_{\alpha}\langle\Psi|F_\alpha|\Psi\rangle^2\nonumber\\ 
       - P \langle\Psi|F_z|\Psi\rangle + Q\langle\Psi|F_z^2|\Psi\rangle,
\end{equation}
where $F_{\alpha=x,y,z}$ is the  spin-1 matrix.
The strength of spin independent interactions is
$c_0=4\pi\hbar^2(a_0+2a_2)/3M>0$, while the 
strength of spin dependent interactions reads
$c_1=4\pi\hbar^2(a_2-a_0)/3M<0$. The constant
$a_S$ is the s-wave scattering length in the total spin $S$
channel ($a_0=101.8a_B$, $a_2=100.4a_B$), and  $M$ is the atom mass. The prefactors of 
linear and quadratic Zeeman shifts are given by 
$P=\mu_B B(\br)/2$ and $Q=\mu_B^2B^2({\bf r})/4E_{hf}$, respectively.
There $E_{hf}$ is the hyperfine splitting energy.

The wave function has three 
condensate components, $\psi_m$,  corresponding to $m=0,\pm1$ projections of
spin-1 onto the magnetic field: $\Psi^T=(\psi_1,\psi_0,\psi_{-1})$. It is
normalized as $\int d{\bf r}\Psi^\dag\Psi= N$, 
where $N$ is the total number of atoms. 
The condensate magnetization reads
$$
f_\alpha= \langle\Psi|F_\alpha|\Psi\rangle, \ \alpha= x,y,z.
$$
It is convenient  to define  a dimensionless parameter 
$$
q(\br)=Q(\br)/n|c_1|,
$$
where $n\approx N/V$ is  the condensate density ($V$ is the system volume).
We consider below setups where $n$ fluctuates negligibly in space.
The critical point corresponds to $q_c=2$.

\section{Quench in a ferromagnetic spin-1 Bose-Einstein condensate}
\label{sec3}
Similarly as in our work on time-dependent quench
\cite{ferro_bodzio_prl,ferro_bodzio_njp}, we assume that 
$\int d\br f_z=0$.
Experimentally, such a constraint can be achieved 
by putting all atoms into an equal mixture of $m=\pm1$ magnetic sublevels 
and letting the sample relax to equilibrium
\cite{leshouches}.
The spin conservation ensures that the final and initial states will have the
same total $f_z$ magnetization: $\int d\br f_z=0$. 
Additionally, we consider  
 $\psi_{\pm1,0}\ge0$ in the ground state: a condition
that can be always imposed because
$
{\cal E}[(|\psi_1|e^{i\chi_1},|\psi_0|e^{i\chi_0},|\psi_{-1}|e^{i\chi_{-1}})]\ge 
{\cal E}[(|\psi_1|,|\psi_0|,|\psi_{-1}|)].
$
This sets $f_y(\br)=0$. 
Moreover, our numerical simulations indicate that  
$f_z(\br)\approx0$ in the spatial quench considered here.
Thus we investigate the $f_x(\br)$ condensate magnetization only.

\begin{figure}[t]
\includegraphics[width=.5\columnwidth,clip=true]{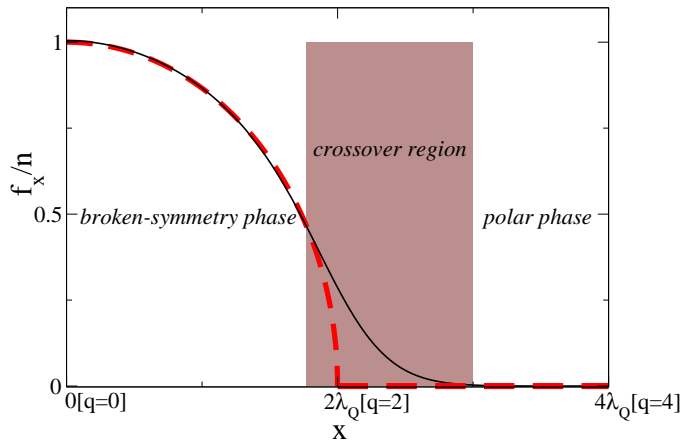}
\caption{A typical picture of the condensate magnetization in a spatial
quench (the black line). The dashed red line shows the magnetization in the
local homogeneous approximation, where the healing length is assumed
to be negligibly small compared to the length scale of variations imposed on the 
condensate. The width of the crossover region 
is proportional to $\hat x$ to  left and right of the spatial 
critical point at $x_c=2\lambda_Q$ [$q(x_c)=2$ (\ref{q_of_r})].
This numerical simulation is for $\lambda_Q/\lambda_0=3/2$ -- see
Appendix \ref{appendix_b} for numerical details.}
\label{smeared}
\end{figure}

For simplicity, we assume that 
\begin{equation}
q(\br)= \frac{x}{\lambda_Q},
\label{q_of_r}
\end{equation}
and forgetting about the gradient term in (\ref{energy})
-- i.e., using our LHA  --
the  ground state   phase diagram can be sketched.
The part of the condensate exposed to  
$0\le q(\br)<2$ is in the ground state of the broken-symmetry phase. There   
$\psi_{\pm1}^{\rm GS}(\br)=\sqrt{n}\sqrt{1/4-q(\br)/8}$ and 
$\psi_0^{\rm GS}(\br)=\sqrt{n}\sqrt{1/2+q(\br)/4}$.
The condensate is magnetized:
$f_x(\br)=n\sqrt{1-q(\br)^2/4}$, while $f_{y,z}(\br)=0$. This ground state breaks 
rotational symmetry on the $(x,y)$ plane  present in  
energy functional (\ref{energy}).
Parts of the condensate exposed to $q(\br)>2$ are 
in the polar phase ground state  with  
$\psi_{\pm1}^{\rm GS}(\br)=0$ and $\psi_0^{\rm GS}(\br)=\sqrt{n}$,
which implies $f_{x,y,z}(\br)=0$.
The dependence of the condensate magnetization
on $x$ and $q$ in the LHA  
is depicted with the red dashed line on Fig. \ref{smeared}. There 
one condensate accommodates the two phases.

The inclusion of the gradient term in the energy functional (\ref{energy}) 
prohibits singularities in the first 
derivative of the wave function. The smooth crossover region replaces 
the sharp phase boundary predicted by the LHA  (Fig. \ref{smeared}).
Its size is proportional to $\hat x$ (\ref{x_hat}), as will be carefully 
discussed below.

In the simplest setup $q(\br)$ dependence  (\ref{q_of_r}) 
is achieved by using the  static, inhomogeneous, magnetic field 
to drive the system between the two phases.
That would correspond to the following 
spatial variation of the Zeeman coefficients
across a typical  $^{87}$Rb condensate: $\Delta Q \sim n |c_1|$ and 
$\Delta P\sim10^4 n|c_1|$. The latter result is obtained 
by using $P=h\times70\times10^4\times B$ HzG$^{-1}$, 
$Q=h\times70B^2$ HzG$^{-2}$ \cite{nature_kurn}, and
$n|c_1|/h= 9.77$ (evaluated at a peak  density 
of the Berkeley experiment \cite{nature_kurn}).
As can be found numerically, such an abrupt $P$-variation 
makes the LHA  phase diagram qualitatively 
incorrect.

To avoid these complications we propose to 
expose the condensate to the inhomogeneous 
magnetic field $B$ pointing in the $z$-direction and  changing 
in time faster than the characteristic time scales 
for the condensate dynamics. When $\langle B(x,t)\rangle_t=0$,
where $\langle\cdots\rangle_t$ denotes time average, 
the linear Zeeman shift will be 
wiped out (which we assume from now on), 
while the quadratic term will be position-dependent only: 
$Q(x)\sim \langle B(x,t)^2\rangle_t$. Thus, we are still dealing 
with a phase transition in space.

To make use of the general predictions worked out in (\ref{x_hat}),
we have to determine the value of the critical exponent $\nu$
and the prefactor $\xi_0$ from  the generic expression for the healing length
(\ref{heal}). We do it by comparing (\ref{heal}) to  the expression for
a divergent  healing length near the critical point (\ref{healing_l}), 
which we have derived in Appendix \ref{appendix_a}. 
We get that $\nu=1/2$
on both sides of the QPT. Introducing a constant 
\begin{equation}
\xi_s=\sqrt{\hbar^2/2Mn|c_1|},
\label{xi_es}
\end{equation}
we also see from (\ref{healing_l}) that $\xi_0$ equals $\xi_s$ in the polar phase
and $\xi_s/\sqrt{2-2|c_1|/c_0}$ in the broken-symmetry phase.

\begin{figure}[t]
\includegraphics[width=.5\columnwidth,clip=true]{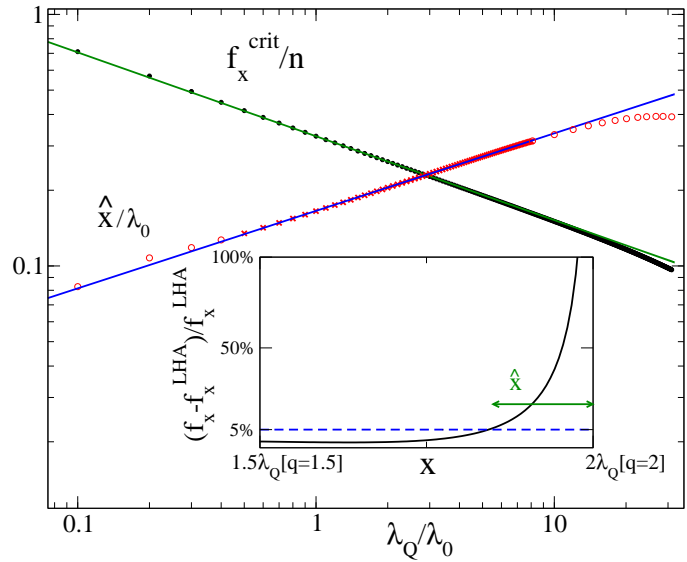}
\caption{Illustration of various scalings.
The black dots: numerics showing the condensate magnetization at the spatial critical point
$f_x^{\rm crit}=f_x(x_c=2\lambda_Q)$. The green line presents
the curve $\sim\lambda_Q^{-1/3}$. 
The red crosses and circles show numerics for
$\hat x$ on the broken-symmetry side of the QPT. 
The blue line presents power law $\exp(-1.7984)(\lambda_Q/\lambda_0)^{0.3082}$,
that was obtained from the fit 
$\ln\hat x/\lambda_0 = -1.7984\pm0.0005 + (0.3082\pm0.0004)\ln\lambda_Q/\lambda_0$, where
$\lambda_0$ is a unit of length (see Appendix \ref{appendix_b}).
The fit was done to data in the range of $\lambda_Q/\lambda_0\in(1/2,8)$ -- 
red crosses. Red circles show numerical data that was not used for this fit.
The determination of $\hat x$ is illustrated on the inset showing the data
for $\lambda_Q/\lambda_0=3/2$: $\hat x$ is the distance from the critical point 
where the relative departures of the condensate magnetization from 
the LHA result, $f_x^{\rm LHA}=n\sqrt{1-q(x)^2/4}$,
start to exceed $10\%$ 
(other thresholds, $1\%-10\%$, give a similar result). Note logarithmic scale on
both axis.}
\label{plot_scaling}
\end{figure}

Coming back to (\ref{x_hat}) we get that the size of the crossover region 
on either the polar or the broken-symmetry side is
\begin{equation}
\hat x=\xi_0^{2/3}\lambda_Q^{1/3}.
\label{scaling}
\end{equation}
This expression can be verified by looking at  experimentally measurable quantity: 
the ground state magnetization of the condensate \cite{nature_kurn}.

We start our discussion on the polar side and linearize the
wave-function around polar phase ground state: $\psi_0=\sqrt{n}(1+\delta\psi_0)$,
$\psi_{\pm1}=\sqrt{n}\delta\psi_{\pm1}$, with $\delta\psi_m\ll1$. This leads to 
$f_x=\sqrt{2}n(\delta\psi_1+\delta\psi_{-1})+O(\delta\psi_m^2)$. 
Using position-dependent $q$ from (\ref{q_of_r}) and linearized version
of (\ref{meanfield}) we arrive at  
\begin{equation}
\xi_s^2\vec{\nabla}^2 f_x = \left(\frac{x}{\lambda_Q}-2\right)  f_x,
\label{dfx}
\end{equation}
which is solved by   
\begin{equation}
f_x(\br) = f_x^{\rm crit} {\rm Ai}\left(\Delta x/(\lambda_Q\xi_s^2)^{1/3}\right)/{\rm Ai}(0).
\label{ai}
\end{equation}
There  $f_x^{\rm crit}$ is the condensate magnetization at the 
spatial critical point (it depends on $\lambda_Q$ only), 
$\Delta x=x-2\lambda_Q\ge0$ is the
distance from the spatial critical point, and $\rm Ai$ is an Airy
function -- the only nondivergent solution of (\ref{dfx}). 
This solution  rigorously
shows that decay of the magnetization, $f_x(\br)/f_x^{\rm crit}$, 
takes place on a length scale $(\lambda_Q\xi_s^2)^{1/3}$ in
full agreement with the strikingly simple scaling result (\ref{scaling}).
Therefore, in the polar phase, 
the magnetization approaches the LHA value, 
$f_x=0$, at a distance $\hat x$ from the critical point.

On the broken-symmetry side of the QPT,
we refer to  numerics (Appendix \ref{appendix_b}) to verify  accuracy of (\ref{scaling}).
In the two extreme limits, very large and very small $\lambda_Q$
the theory does not work well. 
Indeed, using (\ref{q_of_r}) we see that the spatial extent 
of the broken-symmetry phase is $2\lambda_Q$ (Fig. \ref{smeared}). Thus, we expect that 
$\hat x \ll 2\lambda_Q$, i.e., $\lambda_Q\gg\xi_0/2^{3/2}$, 
to see well the crossover region on the broken-symmetry side. 
In the other limit, large $\lambda_Q$, we have to take into account 
that the system size in numerical simulations is finite, say $l$.
There we need to have $\lambda_Q \ll l/2$ so that 
there will be a spatial point in our system where $q(x)=2$, 
and the finite size corrections coming from proximity 
of the system boundary to the spatial critical point 
will be negligible.
These two estimates tell us that $\lambda_Q$ in our system shall be 
much larger than $0.1\lambda_0$ and much smaller than $39\lambda_0$ (see
Appendix \ref{appendix_b} for numerical value of $l$, 
$\xi_0\approx\xi_s/\sqrt{2}$ and the unit of length $\lambda_0$
relevant for our simulations). 
Increasing (decreasing) the lower (upper) limit on $\lambda_Q$ by 
a factor of five we fit a power law to $\lambda_Q/\lambda_0\in(1/2,8)$.
We get from a fit that 
$\ln\hat x/\lambda_0 = -1.7984\pm0.0005 + (0.3082\pm0.0004)\ln\lambda_Q/\lambda_0$ -- see
Fig. \ref{plot_scaling}.
The numerical data for $\hat x$ clearly departs from the fitted line 
for $\lambda_Q/\lambda_0 \gtrsim 10$. Departures for $\lambda_Q/\lambda_0 \lesssim
1/2$ are much less pronounced.
Therefore, we get approximately that $\hat x \sim \lambda_Q^{0.31}$, which 
is in good qualitative agreement with the predicted scaling, i.e., $\lambda_Q^{1/3}$.
We expect that  better agreement can be obtained for large systems
where we can explore larger $\lambda_Q$'s for which 
crossover region is closer to the critical point ($\hat x/\lambda_Q\sim\lambda_Q^{-1/3}$).
More precisely,  by taking the limit of $N,l,\lambda_Q\to\infty$ at 
$\xi_0={\rm const}$ (i.e., $N/l={\rm const}$)  we expect that the scaling (\ref{scaling}) will be
fully  recovered from numerical simulations.

Now let's focus on the magnetization at the critical point.
From (\ref{ai}) we know that it scales in the same way 
as the condensate magnetization at the border of the 
crossover region in the polar phase ($x_+=2\lambda_Q+\hat x$):
$$
f_x(x_+)/n = f_x^{\rm crit}{\rm Ai}(1)/{\rm Ai(0)} \sim f_x^{\rm crit}.
$$
Assuming that similar scaling relation will also hold between $f_x^{\rm crit}$ 
and $f_x$ at $x_-=2\lambda_Q-\hat x$, i.e., 
at the broken-symmetry phase border of the crossover region where 
$$
f_x(x_-)/n \approx\sqrt{1-q(x_-)^2/4}
\approx \sqrt{\hat x/\lambda_Q}\sim \lambda_Q^{-1/3},
$$
we get that $f_x^{\rm crit}\sim\lambda_Q^{-1/3}$.
As depicted in Fig. \ref{plot_scaling}, we have indeed a robust 
$f_x^{\rm crit}\sim\lambda_Q^{-1/3}$ scaling 
for $\lambda_Q/\lambda_0$ in the large range of $10^{-1}$ to $10$.
It is quite remarkable that 
the above intuitive prediction of the  
$f_x^{\rm crit}$ scaling matches numerics so well. It suggests that magnetization 
at the spatial critical point might be a robust observable for
studies of spatial quenches. Similar observation was made for 
condensate magnetization at the critical point during 
a temporal quench  from the broken-symmetry to the polar phase 
\cite{ferro_bodzio_njp}.

All these results are in qualitative agreement 
with our expectations: in  the limit of large $\lambda_Q$ -- very slow
spatial driving -- the condensate ground state approaches  
the LHA. Indeed, both the size of the crossover region
in the $q$-parameter space,
$\hat x/\lambda_Q$
,
and the condensate magnetization at the critical point, $f_x^{\rm
crit}$
, go to zero as $\lambda_Q$ increases.

The measurements of the dependence of the size of crossover region  
on $\lambda_Q$ should allow for the 
first  experimental determination of the scaling exponent 
$\nu$ in a ferromagnetic spin-1 Bose-Einstein condensate. 
Thus, our considerations lead to the new way of investigating the critical region 
of phase transitions. Another scheme that can be used to determine
the exponent $\nu$ was explored experimentally by Esslinger's group in the
context of the classical phase transition from a normal gas to a 
Bose-Einstein condensate \cite{esslinger}. The setup 
involved in this experiment is quite complicated as it requires presence of
a high finesse optical cavity for detection purposes. We expect that the
approach proposed by us should be easier to implement in the 
context of quantum phase transitions.
Moreover, according to (\ref{x_hat}) 
our scheme shall be also ready for experimental exploration 
in the whole zoo of other 
physical systems undergoing a QPT. Finally, we would like to stress that
applicability of Eq. (\ref{x_hat})  is not limited to mean-field theories:
any experimental departures from the $\lambda_Q^{1/3}$ scaling (apart from finite
size corrections) should indicate 
that mean-field value of the critical exponent $\nu$ does not describe the condensate properly.
In such a case the correct value of the  $\nu$ exponent should be easily extracted from 
experimental data with the help of Eq. (\ref{x_hat}).

\section{Summary}

We have explored physics of the QPT in space in 
a ferromagnetic spin-1 Bose-Einstein condensate \cite{nature_kurn}: the most  
flexible system for  studies of  QPTs in space and time.
Our scaling results explain how  singularities of the critical 
point affect the ground state  magnetization of the condensate. They also 
suggest a new way for  the measurement of the critical exponent $\nu$. 
These  findings are generally applicable to all systems undergoing
a second order QPT. 
The quest for the full understanding
of a spatial quench opens up a prospect of  interdisciplinary studies
similarly as time quenches have done  to date \cite{kibble_today}. 

\section{Acknowledgments}

We acknowledge the support of the U.S. 
Department of Energy through the LANL/LDRD Program.

\appendix
\section{Healing length of a ferromagnetic spin-1 condensate}
\label{appendix_a}
The key ingredient of our theory is the divergent healing length
(\ref{heal}). Below we derive it from mean-field
equations 
\begin{eqnarray}
-\frac{\hbar^2}{2M}\vec{\nabla}^2\psi_{\pm1}
+c_0 (|\psi_0|^2+|\psi_1|^2+|\psi_{-1}|^2)\psi_{\pm1}
 + c_1
(|\psi_0|^2+|\psi_{\pm1}|^2-|\psi_{\mp1}|^2)\psi_{\pm1}
+c_1\psi_0^2\psi_{\mp1}^*+Q\psi_{\pm1}=\mu\psi_{\pm1}, \nonumber\\
-\frac{\hbar^2}{2M}\vec{\nabla}^2\psi_0
+c_0 (|\psi_0|^2+|\psi_1|^2+|\psi_{-1}|^2)
\psi_0 + c_1 (|\psi_1|^2+|\psi_{-1}|^2)\psi_0
+2c_1\psi_0^*\psi_1\psi_{-1}=\mu\psi_0,
\label{meanfield}
\end{eqnarray}
that come from minimization of ${\cal E}-\mu\int d{\br}\sum_m|\psi_m|^2$
without the linear Zeeman term (\ref{energy})
-- see Sec. \ref{sec3} for the explanation of why we remove it from our considerations. 

The healing length is a typical length scale 
over which a local perturbation of the wave-function gets
forgotten (here we will have three characteristic length scales, but we will identify
the leading one relevant for a long distance healing process). To find it, 
we assume the constant  parameter $q=Q/n|c_1|$ across the condensate,
and linearize  mean-field equations (\ref{meanfield}).
We write the wave-function as 
$\psi_{m}=\psi_m^{\rm GS}+\sqrt{n}\delta\psi_m$ with 
$m=0,\pm1$ and $\psi_m^{\rm GS}$ being the ground state solution. 
In the broken-symmetry phase we have 
$\psi_{\pm1}^{\rm GS}(\br)=\sqrt{n}\sqrt{1/4-q(\br)/8}$ and 
$\psi_0^{\rm GS}(\br)=\sqrt{n}\sqrt{1/2+q(\br)/4}$, while 
in  the polar phase,  
$\psi_{\pm1}^{\rm GS}(\br)=0$ and $\psi_0^{\rm GS}(\br)=\sqrt{n}$.
We use here the freedom to work with $\psi_{0,\pm1}\ge0$, which is
explained in Sec. \ref{sec3}.

Linearized equations take the form
\begin{equation}
\xi_s^2\vec{\nabla}^2\delta\psi_m=\sum_k S_{mk} \delta\psi_k, 
\label{S_eq}
\end{equation}
where $\xi_s=\sqrt{\hbar^2/2Mn|c_1|}$ and  $S$ stands for 
$$
\frac{c_0}{4|c_1|}\left(
\begin{array}{ccc}
2-qu_-      &    (4+2q)u_- &  2-qu_+  \\
(2-q)u_-    &    4+2q      &  (2-q)u_-\\
2-qu_+      &    (4+2q)u_- &  2-qu_-
\end{array}
\right), \ u_\pm = 1\pm2|c_1|/c_0
$$
in the broken-symmetry phase, and for 
$$
\left(
\begin{array}{ccc}
q-1  &   0            &  -1 \\
0    &   2c_0/|c_1|    &  0  \\
-1   &    0           &  q-1
\end{array}
\right)
$$
in the polar phase. The factor $u_\pm$ above is very close to unity as 
$|c_1|/c_0\approx1/216.1\ll1$ for $^{87}$Rb atoms considered here.

To proceed we diagonalize matrix $S$ by the transformation $S=C D C^{-1}$,
where $D={\rm diag}(\Gamma_1,\Gamma_2, \Gamma_3)$ is a diagonal matrix 
made of eigenvalues of matrix $S$. In the polar phase we find
$$
\Gamma_m = q-2,q,2c_0/|c_1|,
$$
while in the broken-symmetry phase we get 
$$
\Gamma_m = c_0/|c_1|\mp\sqrt{(c_0/c_1)^2-(4-q^2)(c_0/|c_1|-1)}, q.
$$

After defining $\delta\tilde\psi_m = \sum_n
C^{-1}_{mn}\delta\psi_n$ and some elementary algebra we get
$$
\frac{\xi_s^2}{\Gamma_m} \vec{\nabla}^2\delta\tilde\psi_m =
\delta\tilde\psi_m \Rightarrow \delta\tilde\psi_m = f\left(\frac{\br}{\xi_s/\sqrt{\Gamma_m}}\right),
$$
where the form of the function $f$ depends on system dimensionality (1D, 2D,
or 3D), while $\br$ is ``attached'' to the point in space where the perturbation is imposed 
on the wave-function. This solution implies that 
the eigenvalue vanishing at the critical point,  $\Gamma_{\rm min}$, provides the leading contribution 
to the long-distance healing process near the critical point ($q\approx2$). 
Indeed, a simple calculation shows that the wave function will forget about 
the perturbation imposed on it
($\delta\psi_m,\delta\tilde\psi_m\approx 0$)
at a distance scaling as $\xi_s/\sqrt{\Gamma_{\rm min}}$.
Thus, the divergent healing length equals $\xi_s/\sqrt{\Gamma_{\rm min}}$. 
Around the critical point 
we find it to be  
\begin{equation}
\xi^{\rm broken} 
=\frac{\xi_s}{\sqrt{2-2|c_1|/c_0}|q-2|^{1/2}}, \ \
\xi^{\rm polar}=\frac{\xi_s}{|q-2|^{1/2}}
\label{healing_l}
\end{equation}
on the broken-symmetry ($q<2$) and  polar ($q>2$) sides
of the QPT, respectively. The expansion around $q=2$ was applied to 
obtain $\xi^{\rm broken}$, i.e., we have approximated the eigenvalue
$c_0/|c_1|-\sqrt{(c_0/c_1)^2-(4-q^2)(c_0/|c_1|-1)}$ by 
$(2-2|c_1|/c_0)(2-q)$ near the critical point.
These expressions fix the value of the mean-field critical exponent $\nu$ for our 
model: $\nu=1/2$ (compare (\ref{healing_l}) to (\ref{heal})).
We conclude by showing another simplification of the full expression 
for the divergent healing length on the broken symmetry side. Taking the limit of
$|c_1|/c_0\to0$, well justified for $^{87}$Rb, we get 
$$
\xi^{\rm broken} = \frac{\xi_s\sqrt{2}}{\sqrt{4-q^2}}, 
$$
without restricting ourselves to $q\approx2$.

\section{Details of numerical simulations}
\label{appendix_b}
Our analytical predictions are valid in any dimensional system, but 
we test them numerically in a 1D model. 
Our numerics is based on  conjugate gradient minimization of the
1D energy functional 
\begin{equation}
     \int d\tilde x \ 
         \frac{1}{2}\frac{d\tilde \Psi^T}{d\tilde x}\frac{d\tilde \Psi}{d\tilde x}
       + \frac{\tilde c_0}{2}\langle\tilde\Psi|\tilde\Psi\rangle^2 
       + \frac{\tilde c_1}{2}\sum_{\alpha=x,y,z}\langle\tilde\Psi|F_\alpha|\tilde\Psi\rangle^2\nonumber\\ 
       + q \tilde n|\tilde c_1|\langle\tilde\Psi|F_z^2|\tilde\Psi\rangle,
\label{energy1D}
\end{equation}
obtained through dimensional 
reduction of the full 3D energy functional (\ref{energy}) without the linear
Zeeman term 
-- see Sec. \ref{sec3} for the explanation of why we remove it from our considerations. 
Moreover, all the components of the above energy functional are dimensionless, and
chosen to match qualitatively the experiment of the Berkeley
group \cite{nature_kurn}. Derivation of (\ref{energy1D}) is explained in detail 
in \cite{ferro_bodzio_prl}: please account for different normalization of the wave-function 
between this paper and \cite{ferro_bodzio_prl}.

In short,  we have $N$, number of atoms, equal to 
$2\times10^6$, $N\tilde c_1= -629$, and $\tilde c_0=216.1 |\tilde c_1|$.
The unit of length used through the paper, $\lambda_0$, equals $5.1\mu m$. 
The system (box) size is $\tilde l = l/\lambda_0 = 78$, i.e., about $0.4 mm$ 
($\tilde l\times\lambda_0$).
The coordinate $\tilde x$ runs from $0$ to $\tilde l$. 
The density, $\tilde n$, equals $N/\tilde l$,
while $\int d\tilde x \tilde\Psi^T\tilde\Psi=N$. 
The ``prefactor'' of the healing length, $\xi_s$ (\ref{xi_es}), in these dimensionless 
quantities is $\sqrt{\tilde l/2N|\tilde c_1|}\approx 1/4$, i.e., about $1.3\mu m$ 
($\lambda_0/4$).


\begin{thebibliography}{99}
\bibitem{healing} The healing length is the typical   distance  over which a
                   system can adjust to a local perturbation imposed on it.
\end{thebibliography}
\end{document}